\documentclass[times,twocolumn,final,numbers]{elsarticle} 
\usepackage{array}
\usepackage{subfig}  
\usepackage{multirow}
\usepackage{hhline}
\usepackage{mathtools} 
\usepackage{gensymb}   
\usepackage{amsmath}                   
\usepackage{amssymb}
\usepackage{mathptmx}
\usepackage{arydshln}
\usepackage{comment}
\usepackage[hidelinks]{hyperref} 
\usepackage{xcolor}
\usepackage{algorithm}
\usepackage{algpseudocode}

\usepackage{color}
\usepackage{amsmath}

\def\fixedlabel#1#2{
	\@bsphack
	\protected@write\@auxout{}
	{\string\newlabel{#1}{{#2}{\thepage}}}
	\@esphack}

\newcommand{\ie}{\textit{i.e.}}
\newcommand{\eg}{\textit{e.g.}}

\journal{Medical Engineering \& Physics}

\begin{document} 

\begin{frontmatter}

\title{Bone Surface Reconstruction and Clinical Features Estimation from Sparse Landmarks and Statistical Shape Models: A feasibility study on the femur}

\author[myadd,ltadd]{Alireza Asvadi\corref{mycorrespondingauthor}}
\cortext[mycorrespondingauthor]{Corresponding author at: UFR M{\'e}decine 22, avenue Camille Desmoulins C.S. 93837, 29238 Brest Cedex 3, France.} 
\ead{alireza.asvadi@gmail.com}

\author[gdadd,ltadd]{Guillaume Dardenne}
\author[jtadd]{Jocelyne Troccaz}
\author[vbadd,ltadd]{Val{\'e}rie Burdin}

\address[myadd]{University of Western Brittany, UBO, Brest France;}
\address[gdadd]{University Hospital of Brest, Brest, France;}
\address[jtadd]{Univ. Grenoble Alpes, CNRS, Grenoble INP, TIMC-IMAG, F-38000 Grenoble, France;}
\address[vbadd]{IMT Atlantique, Mines Telecom Institute, Brest, France;}
\address[ltadd]{Laboratory of Medical Information Processing (LaTIM),  INSERM U 1101, Brest, France;} 

\begin{abstract}
In this study, we investigated a method allowing the determination of the femur bone surface as well as its mechanical axis from some easy-to-identify bony landmarks. The reconstruction of the whole femur is therefore performed from these landmarks using a Statistical Shape Model (SSM). 
The aim of this research is therefore to assess the impact of the number, the position, and the accuracy of the landmarks for the reconstruction of the femur and the determination of its related mechanical axis, an important clinical parameter to consider for the lower limb analysis. Two statistical femur models were created from our in-house dataset and a publicly available dataset. Both were evaluated in terms of average point-to-point surface distance error and through the mechanical axis of the femur. Furthermore, the clinical impact of using landmarks on the skin in replacement of bony landmarks is investigated.
The predicted proximal femurs from bony landmarks were more accurate compared to on-skin landmarks while both had less than $3.5 ^{\circ}$ degrees mechanical axis angle deviation error. 
The results regarding the non-invasive determination of the mechanical axis are very encouraging and could open very interesting clinical perspectives for the analysis of the lower limb either for orthopedics or functional rehabilitation.
\end{abstract}


\end{frontmatter} 


\section{Introduction} \label{sec:introduction} 
Statistical models are increasingly recognized in medical image analysis and particularly in our domain of interest: computer-assisted orthopedic surgery \cite{joskowicz2018future}. 
The Statistical Shape Model (SSM) enables an efficient parameterized representation of the shape even when only a few shape instances are available \cite{cootes1992training, rueckert2019model}.
It is not only useful for the compact modeling of the anatomical shape variation in the dataset but also to generate new shapes similar to the original data and to reconstruct missing parts of shapes \cite{bernard2017shape, salhi2020statistical}.

Shape reconstruction is usually performed using a SSM and some constraints which encompass the difference between the SSM and the intact part of the shape.
These constraints will guide the deformation of the SSM and fit it to the partial present part of the shape and predict the missing part.
Several approaches have been proposed in this context, mainly different in the way of building the SSM or in the type of constraints used.
Zhang and Besier \cite{zhang2017accuracy} and Blanc et al. \cite{blanc2012statistical} used sparse observations and morphometric data as the constraints.
Bernard et al. \cite{bernard2017shape} used sparse point data and employed Gaussian mixture models (GMM) to adjust the SSM.
Rajamani et al. \cite{rajamani2007statistical} performed reconstruction using a few landmarks and surface points and a least-squares minimization approach. Albrecht et al. \cite{albrecht2013posterior} introduced posterior SSM to predict the best matching shape.

\begin{figure} [h!]
	\centering
	\includegraphics[width=6.4cm]{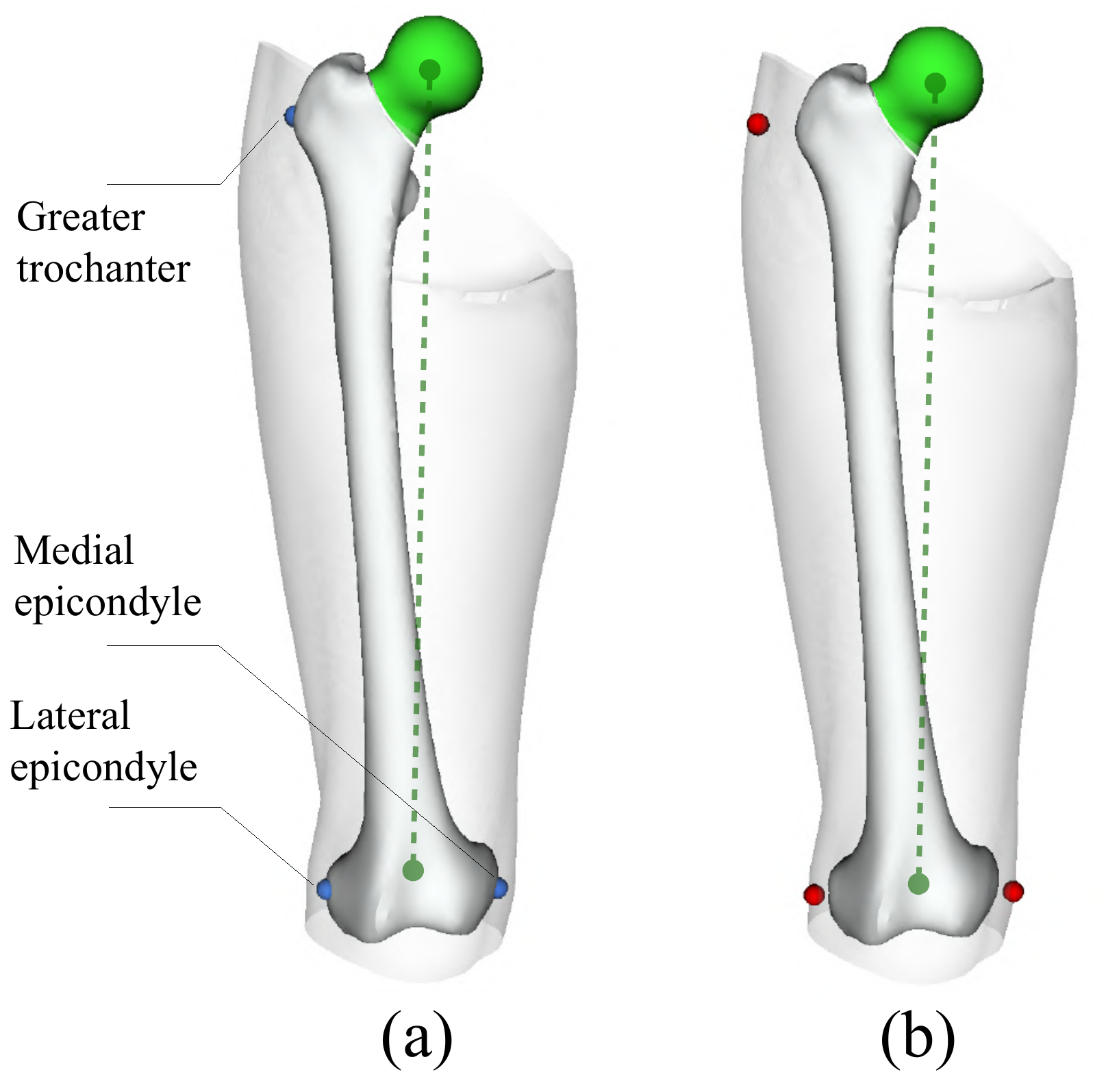}
	\caption{Our main objectives are the reconstruction of the missing proximal femur (shown in green) and estimation of the mechanical axis (dashed green line) using only three landmarks: (a) bony landmarks (indicated in blue) and (b) on-skin palpation guidance (shown in red).}
	\label{figure_pipeline_abs2} 
\end{figure}

While most of these works focus on developing new algorithms for shape prediction, an issue that has not been fully investigated is the impact of the quality of the constraints on the reconstruction performance. 
In this study, we assessed the impact of the position, the noise, and the number of easily reachable bony landmarks (as the constraints to guide the deformations) for the non-invasive reconstruction of a femur, and for the estimation of its mechanical axis: the axis between the hip and the knee centers. 
Several strategies regarding the acquisition of landmarks have been investigated and compared, either landmark directly digitized on the bone surface such as for instance, during knee surgery or through the skin, \ie, in a non-invasive way (see Fig. \ref{figure_pipeline_abs2}). 

These findings could have great importance either for orthopedics or rehabilitation, to the further non-invasive estimation of clinical parameters which allows an optimal assistance for the placement of knee implant, or for improving the preoperative or postoperative patient's follow-up.
The reconstruction of the femur bone from partial data allows the determination of specific lower limb parameters such as the hip joint center or the mechanical axis. 
The mechanical axis is a crucial clinical indicator for lower limb alignment, allowing the assessment of the patient biomechanical balance in rehabilitation and orthopedics \cite{zahedi1986alignment}.
The precise alignment of the lower limb also has a crucial role in knee joint-related surgeries \cite{rahm2017postoperative, sabzevari2016high}.

\section{Materials and Methods} \label{sec:material}

\subsection{Specimens} \label{sec:specimens} 

\begin{figure} 
\centering
\includegraphics[width=6.5cm]{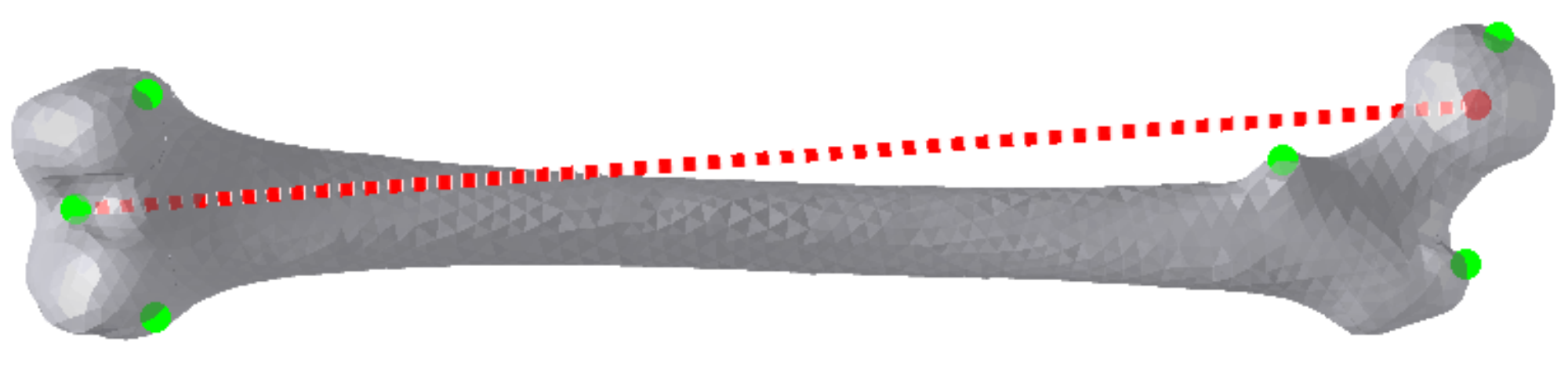}
\caption{The 6 anatomical landmarks (indicated in green) used for the initial rigid alignment. The mechanical axis is shown in red dashed line. }  
\label{figure_lms} 
\end{figure}

\noindent \textit{Our femur mesh dataset:}
consists of 20 cadaveric femurs which were scanned at the University Hospital of Brest (CHRU Brest). It is composed of 12 right and 8 left femurs which were mirrored through the sagittal plane to become ``virtually right''. A quadratic decimation process with Meshlab\footnote{\href{http://www.meshlab.net/}{http://www.meshlab.net/}} was performed on each mesh to homogenize and reduce the number of vertices to about $25000$. Six anatomical landmarks were manually located on each femur for the initial rigid alignment purpose (see Fig. \ref{figure_lms}): the fovea, greater trochanter, lesser trochanter, the medial and lateral points of the posterior condyles, and the intercondylar notch. Our dataset was shuffled and partitioned into two groups: $70\%$ training (14 meshes) and $30\%$ test (6 meshes) sets.

\noindent \textit{Sicas femur mesh dataset:}
taken from the Swiss Institute for Computer Assisted Surgery (SICAS) medical image repository\footnote{\href{https://www.smir.ch/}{https://www.smir.ch/}} \cite{kistler2013virtual}. It consists of a set of 50 misaligned femur triangle meshes (on average about $18000$ points per mesh) and their corresponding annotated landmarks (similar to ours). For the analysis, the Sicas femur data was organized into two subsets: $70~\%$ as a training set (35 meshes) and $30~\%$ as a test set (15 meshes).

The femur length indicator (defined as the distance between the fovea and notch landmarks) in our and Sicas datasets were measured as $410\pm26$ and $360\pm24$ mm, respectively. Our data contains femurs with larger lengths whereas length is more evenly distributed in Sicas data (for more details please refer to the supplementary material, section 1).

\noindent \textit{Femur bone and thigh skin mesh data:} 
Five femurs and their enveloping thigh skin mesh data from frozen cadavers were used for the validation study. Mesh data was created from CT scan files using 3D Slicer \cite{Kikinis2014}. The number of vertices in the original femur and skin meshes was decimated to $2000$ and $5000$ vertices, respectively. Figure \ref{figure_pipeline_abs2} illustrates an example of data with the three considered points which are of significant importance in the hip and knee regions and palpated by the clinicians.

\subsection{Creation of Statistical Shape Models} \label{sec:ssm_creation}
The femur shapes were not pre-aligned and the different meshes were not in correspondence. Therefore, a preprocessing step was applied to rigidly align the meshes in both femur datasets using the provided landmarks (see Fig. \ref{figure_lms}) and a Procrustes analysis. Next, the Iterative Median Closest Point (IMCP) \cite{jacq2008performing} - an algorithm for the robust simultaneous rigid registration of all shapes - was used to create the so-called reference unbiased mesh in order to reduce the bias that can occur by a random selection of the reference mesh. The obtained reference mesh was decimated to approximately 5000 points.

One of the essential steps in building statistical shape models is the dense correspondence matching among the meshes. The Coherent Point Drift (CPD) algorithm \cite{myronenko2010point} was used to establish automatic dense correspondences between the training sets and the unbiased mesh. The CPD algorithm is a probabilistic approach for the non-rigid registration of point sets. It treats the unbiased mesh vertices as Gaussian mixture models (GMM) centroids that are fitted in the target mesh vertices by maximizing a probability density function. The MATLAB \textit{`pcregistercpd'} function\footnote{\href{https://www.mathworks.com/help/vision/ref/pcregistercpd.html}{https://www.mathworks.com/help/vision/ref/pcregistercpd.html}} was used for the implementation. The original meshes were replaced by the in-correspondence ones and the in-correspondence training dataset was established (with the same topology and same index of vertices).

The next step was the SSM construction of the femur bone. The Gaussian Process Morphable Models (GPMMs) method \cite{luthi2017gaussian} - a recent algorithm to obtain a continuous model of the shape variability - was adopted to derive the SSM. The open-source SCALISMO framework\footnote{\href{http://github.com/unibas-gravis/scalismo}{http://github.com/unibas-gravis/scalismo}} was used for this implementation.
Two separate SSMs were developed from the Sicas and our train sets using in-correspondence meshes. The created SSMs were used for the proximal femur reconstruction in the next step and their performances were compared. 


\begin{figure*}
\centering
\includegraphics[width=16cm]{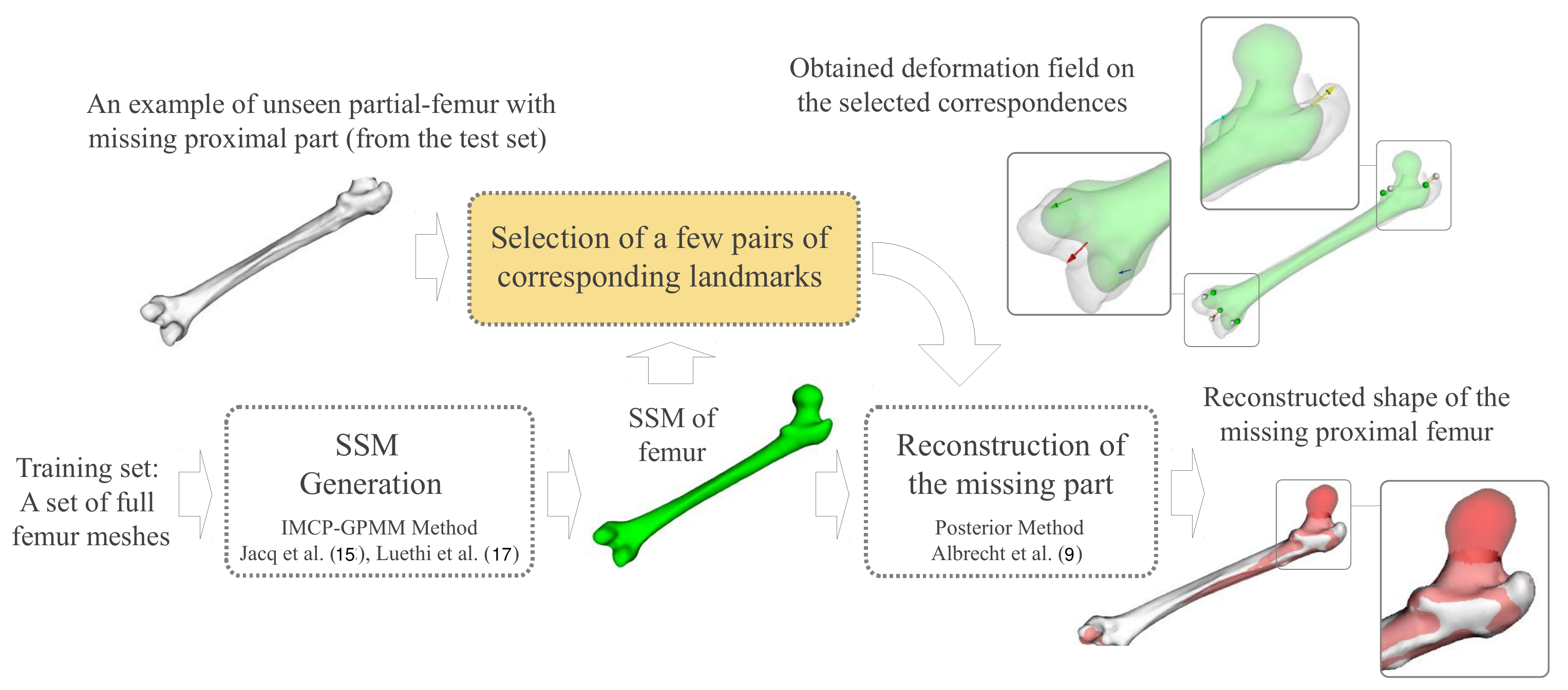}
\caption{The femur reconstruction pipeline. An example of unseen partial femur mesh (in gray), the developed femoral SSM on the training set (in green), and the mean shape in the posterior model (in red) which is obtained using the SSM and the deformation field.} 
\label{figure_pipeline_abs} 
\end{figure*}

\subsection{Reconstruction} \label{subsec:recon} 
The conceptual pipeline of the proximal femur reconstruction from the intact part is summarized in Fig. \ref{figure_pipeline_abs}. The input to the algorithm is an unseen partial femur (from the test set) and the objective is to estimate its missing part. The algorithm can be described by two major steps: development of the femoral SSM; and reconstruction of the missing part using the obtained SSM and a few landmarks as constraints. 
The SSM learnt on the training data was used for the reconstruction of the missing head and neck of the femur bones in the test sets. To predict the missing shape, `posterior shape models' approach \cite{albrecht2013posterior} was used which in fact computes the conditional distribution of a SSM given information about the partial present parts of the shape. To compute the posterior SSM, in addition to the SSM, a deformation field is required. It deforms the SSM to fit the partial femur and to predict the missing part. It is obtained from a pair of source-target landmarks sets. The target landmarks are from the partial present part of the test mesh and the source landmarks are from the SSM reference mesh. The mean shape in the posterior model was considered as the optimal predicted femur. 

One of the main purposes of this study is the estimation of the mechanical axis. A sphere was fitted to 30 random points from the reconstructed femoral head surface using the least square method and the hip center was calculated. The vector from the intercondylar notch to the hip center was considered as the mechanical axis of the femur. 

\subsection{Assessments} \label{subsec:assess} 
\subsubsection{Assessment of landmark configurations}
\begin{figure*} 
\centering
\begin{tabular}{ ccc }
\includegraphics[height=3.5cm]{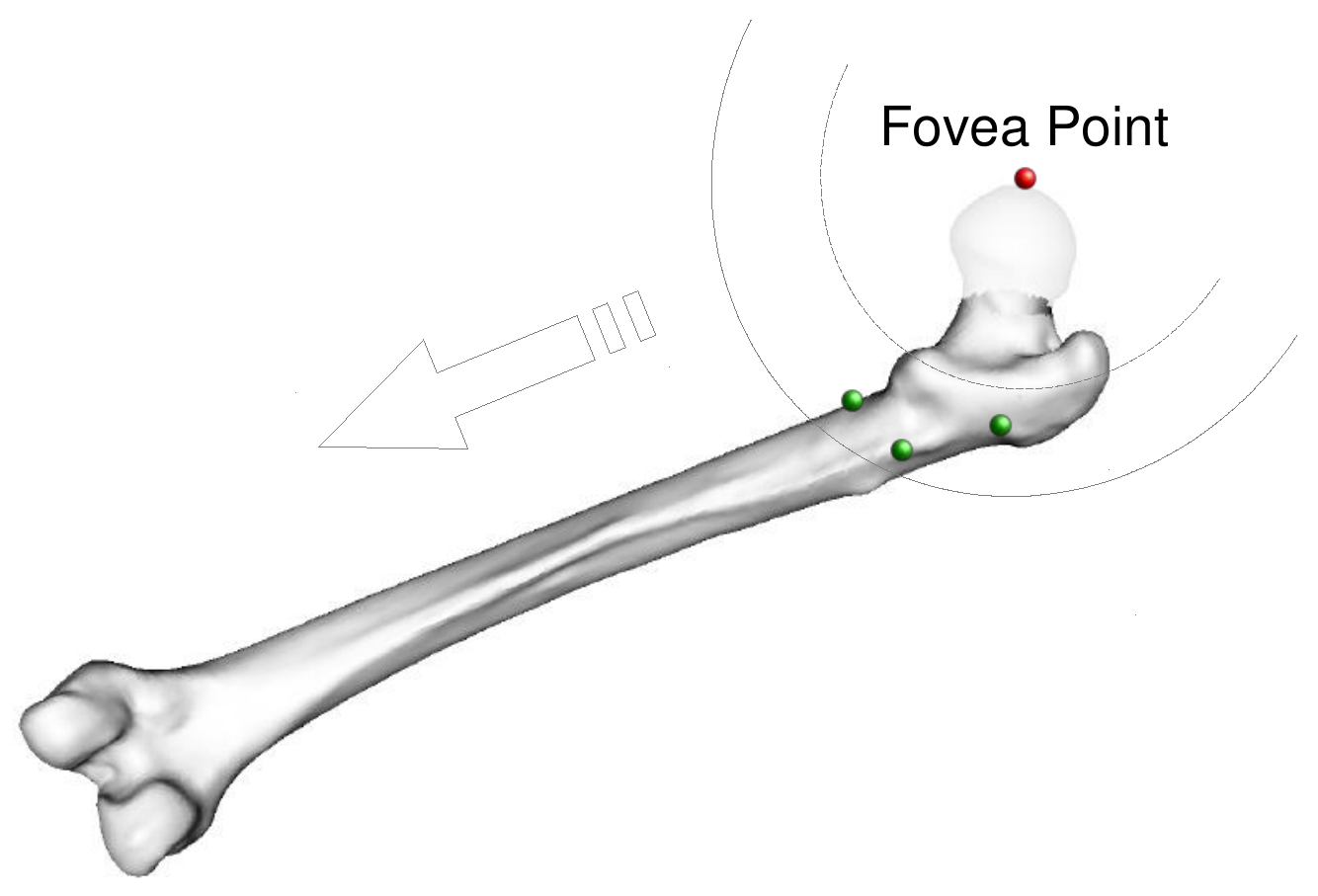} & \includegraphics[height=3.5cm]{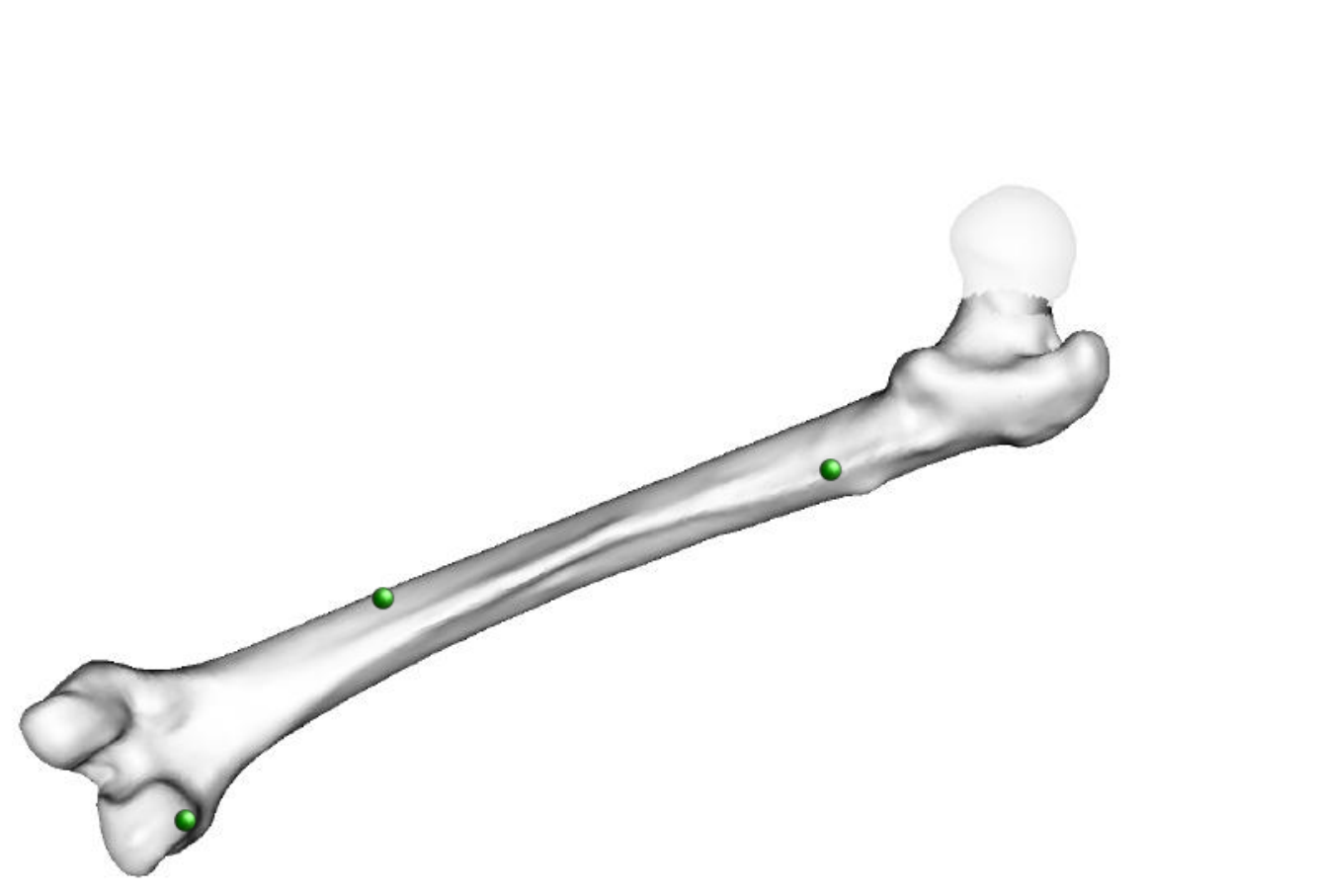} & \includegraphics[height=3.5cm]{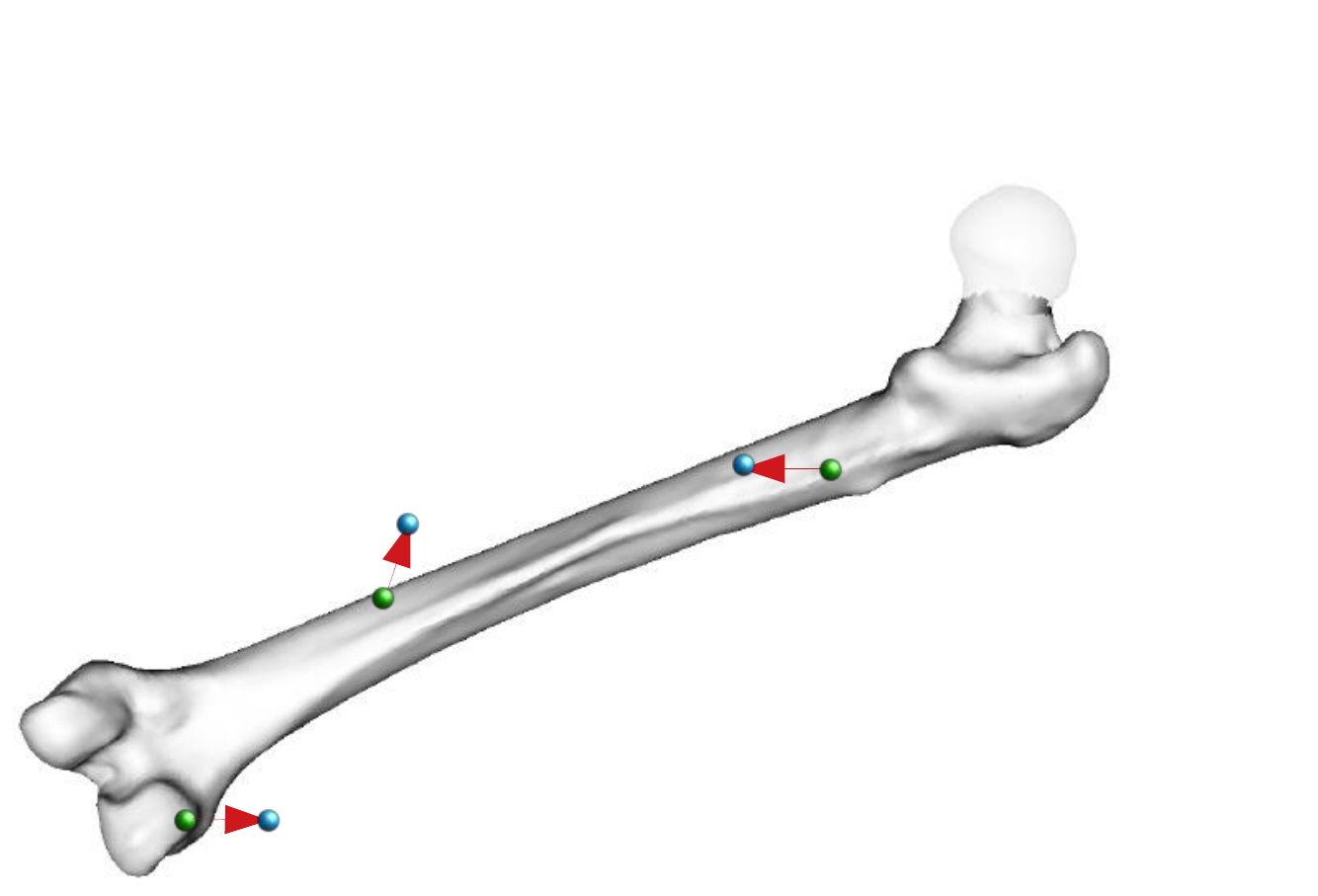} \tabularnewline
\includegraphics[height=3.5cm]{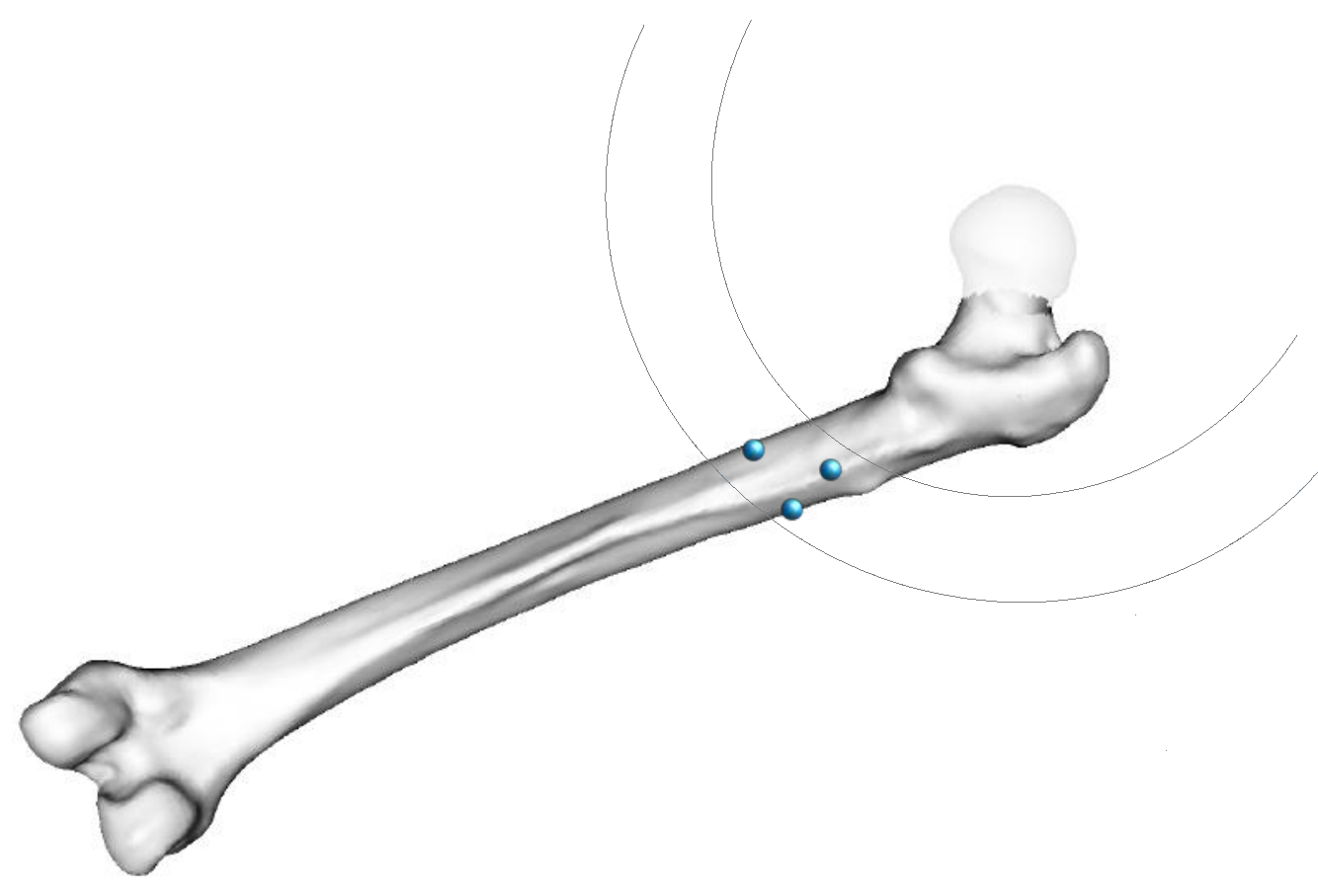} & \includegraphics[height=3.5cm]{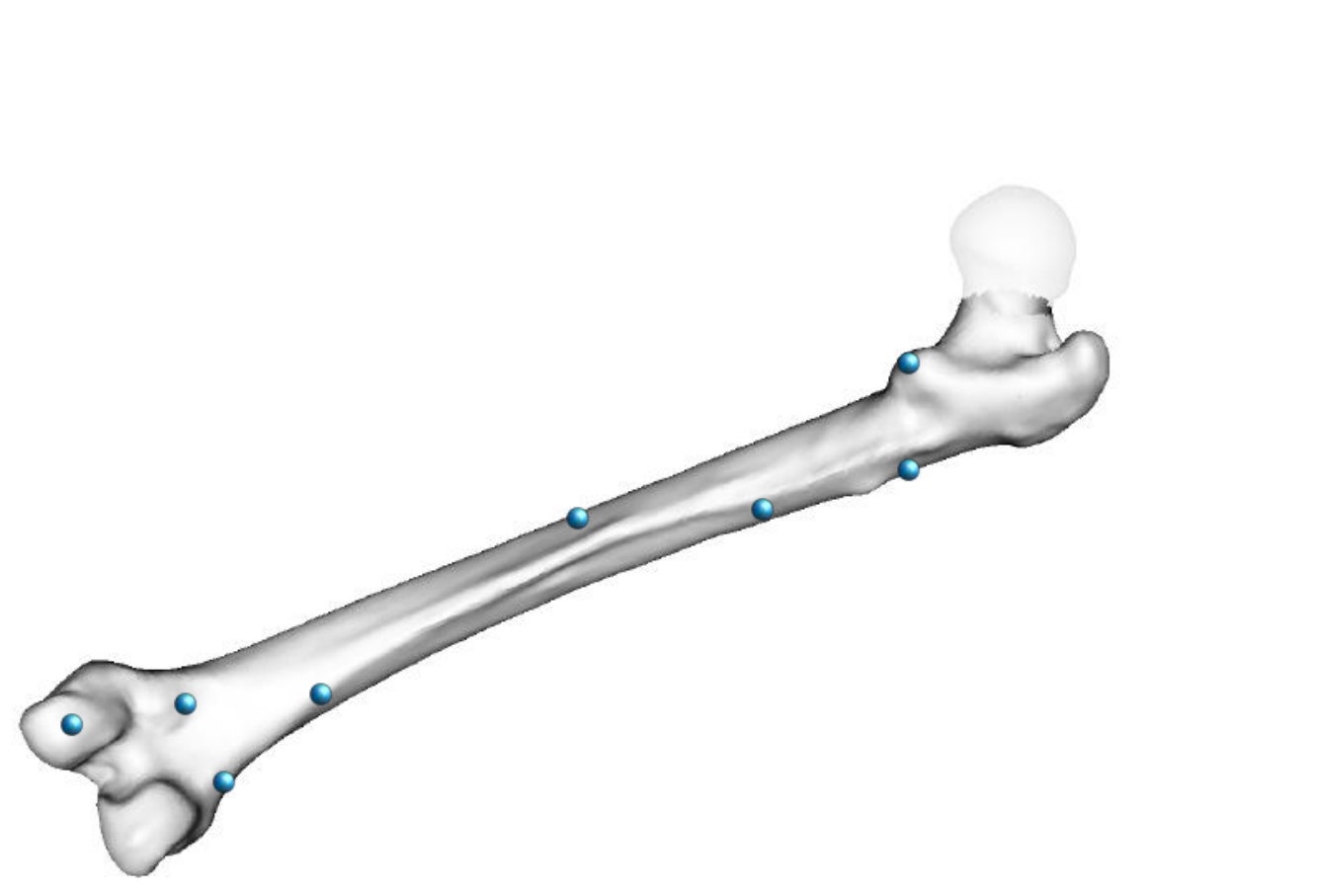} & \includegraphics[height=3.5cm]{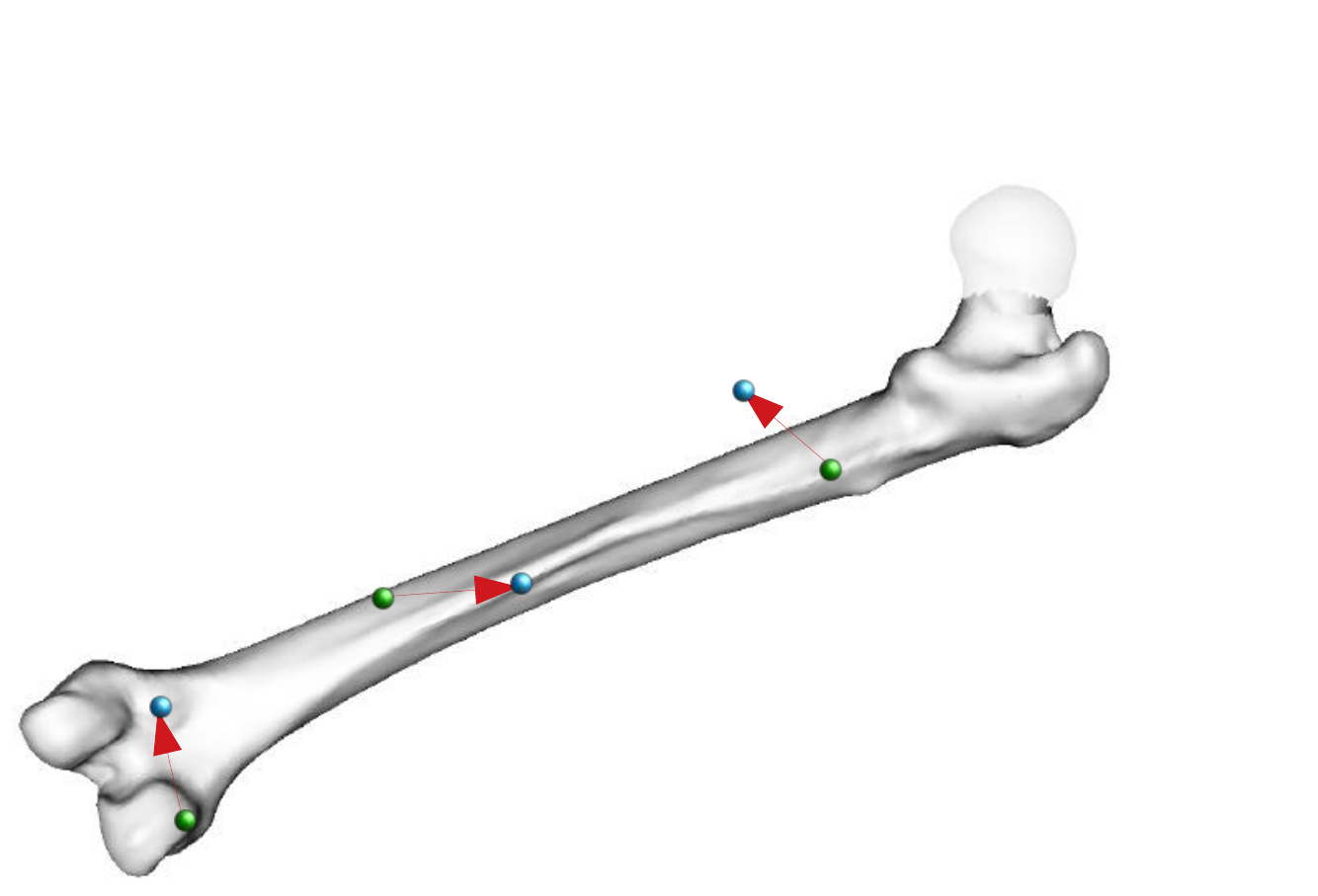} \tabularnewline \tabularnewline
{\centering \footnotesize (a) Distance of landmarks} & 
{\centering \footnotesize (b) $\#$ of landmarks} & 
{\centering \footnotesize (c) Landmarks' displacement} 
\end{tabular}
\caption{Different configuration of landmarks that were used for the evaluation of the reconstruction performance.} 
\label{figure_lmconfig}
\end{figure*}

Experiments were conducted to assess the reconstruction performance when one of the following parameters of the landmarks changes: i) the distance of landmarks from the fovea landmark point, ii) the number of landmarks and iii) inaccuracy in the position of the landmarks. 
In the first experiment, a set of 3 landmarks was considered on the unbiased reference mesh with their corresponding landmarks set on the test mesh. The experiment started with landmarks in the ring-shaped region within $10\%$ and $20\%$ of the `femur length indicator' distance from the fovea landmark. It continued with increasing distance of the region ($20\%$ to $100\%$ of the femur in the step of $10\%$) from the fovea in 8 steps to cover the femur shape. In the left column in Fig. \ref{figure_lmconfig}, top and bottom figures illustrate steps 2 and 3 of this process, covering $20\%$ to $30\%$ and $30\%$ to $40\%$ of the femur respectively. 
The second experiment was performed to evaluate the reconstruction performance when the number of landmarks increases. The middle column in Fig. \ref{figure_lmconfig}, moving from top to bottom, shows this concept. In this experiment, the number of landmarks starts from 5 to the maximum number of 205 landmarks in steps of 50. 
The third experiment was carried out to assess the prediction performance when a bias is induced to the positions of the landmarks on the test mesh (to simulate inaccurate landmark positioning). This case is illustrated in the right column of Fig. \ref{figure_lmconfig}. The landmarks were displaced from the original positions in random directions but with a fixed value. The displacement was made in the range of -5 to +5 mm and steps of 2 mm. 

The reconstruction and assessment were performed in the following steps: clipping the femur mesh into proximal and remaining distal parts (\ie, the femur without the proximal part); keeping the proximal part of the original test mesh for the evaluation and using the remaining distal part for the generation of landmarks. As will be explained later, the remaining distal part of the fitted unbiased mesh to a test mesh using the CPD algorithm was used for the landmark generation (and not that of the original test mesh); computing the posterior model using the landmarks and finally calculating the error between the proximal part of the estimated posterior model and the ground-truth test mesh. The implementation of these steps is explained in the next few paragraphs.

In the first step, the proximal parts $\mathbf{P}$ of the femurs in the test set $\mathbf{T}$ had to be clipped. The clipping process was performed using the fovea (head) landmark point $\mathbf{L}_\mathrm{k}$ of each test mesh $\mathbf{T}_\mathrm{k}$. The vertices of the test mesh with a distance under $10\%$ of the length between the fovea and the furthest node in the femur mesh were considered as the femoral head and neck vertices and kept for the evaluation purpose.
The next step is the automatic correspondence generation between the selected unbiased mesh landmarks $\mathrm{\dddot U}$ and the test mesh landmarks $\mathbf{\dddot T}_\mathrm{k}$. Finding the correspondences and consecutively the vector deformation field $\mathbf{D}_\mathrm{k}$ is necessary for obtaining the posterior model $\mathbf{E}_\mathrm{k}$. To find the correspondences, the process started by fitting the unbiased mesh $\mathrm{U}$ to a test mesh $\mathbf{T}_\mathrm{k}$ using the CPD algorithm. 
The CPD algorithm here was only used to automatically find the equivalent anatomical landmarks from the SSM to the partial shape.
The fitted unbiased mesh to the test mesh $\mathbf{F}_\mathrm{k}$ was then clipped to extract its distal part $\mathbf{B}_\mathrm{k}$. 
The landmark selection was performed by downsampling $\mathbf{B}_\mathrm{k}$ according to one of the experiments described earlier (see Fig. \ref{figure_lmconfig}). 
The downsampled $\mathbf{\dddot B}_\mathrm{k}$ was used as a medium to find the selected pair of unbiased-test landmarks sets $(\mathrm{\dddot U}, \mathbf{\dddot T}_\mathrm{k})$.
The closest points on the test mesh vertices $\mathbf{T}_\mathrm{k}$ to $\mathbf{\dddot B}_\mathrm{k}$ and the corresponding indices in $\mathrm{U}$ were defined as the pair of $\mathrm{\dddot U}$ and $\mathbf{\dddot T}_\mathrm{k}$ landmarks sets.
The deformation field $\mathbf{D}_\mathrm{k}$ was determined from this pair of landmarks sets. The prediction of the missing proximal part was performed by computing the SSM posterior $\mathbf{E}_\mathrm{k}$ guided by the deformation vector field $\mathbf{D}_\mathrm{k}$. The average shape in the posterior model is considered to be the predicted femur shape.

The evaluation was performed in terms of the surface reconstruction error and mechanical axis deviation error. The surface reconstruction error $\mathbf{R}_\mathrm{k}$ between the proximal parts of the ground-truth test mesh $\mathbf{P}_\mathrm{k}$ and the estimated posterior $\mathbf{E}_\mathrm{k}$ was computed in terms of vertex-to-vertex Root Mean Square Error (RMSE). The deviation error $\mathbf{M}_\mathrm{k}$ for each test mesh $\mathbf{T}_\mathrm{k}$ is the 3D angle between the obtained mechanical axis of $\mathbf{\check E}_\mathrm{k}$ and the axis of the ground truth $\mathbf{\check T}_\mathrm{k}$. The experiment was performed in several iterations and the average of errors was computed. The details of the implemented algorithm can be found in the supplementary material, section 2.

\subsubsection{Assessment of skin landmarks}
The objective, at this point, is to study the reconstruction performance when only skin landmarks are used in replacement of bony landmarks. Three landmarks were of our interest as they were considered to be the most accessible for palpation during surgical procedures: greater trochanter, medial epicondyle, and lateral epicondyle (see Fig. \ref{figure_pipeline_abs2}).
Two experiments were set up to examine the effect of soft-tissue artifacts. 

\begin{figure} [t!]
\centering
\includegraphics[width=6.5cm]{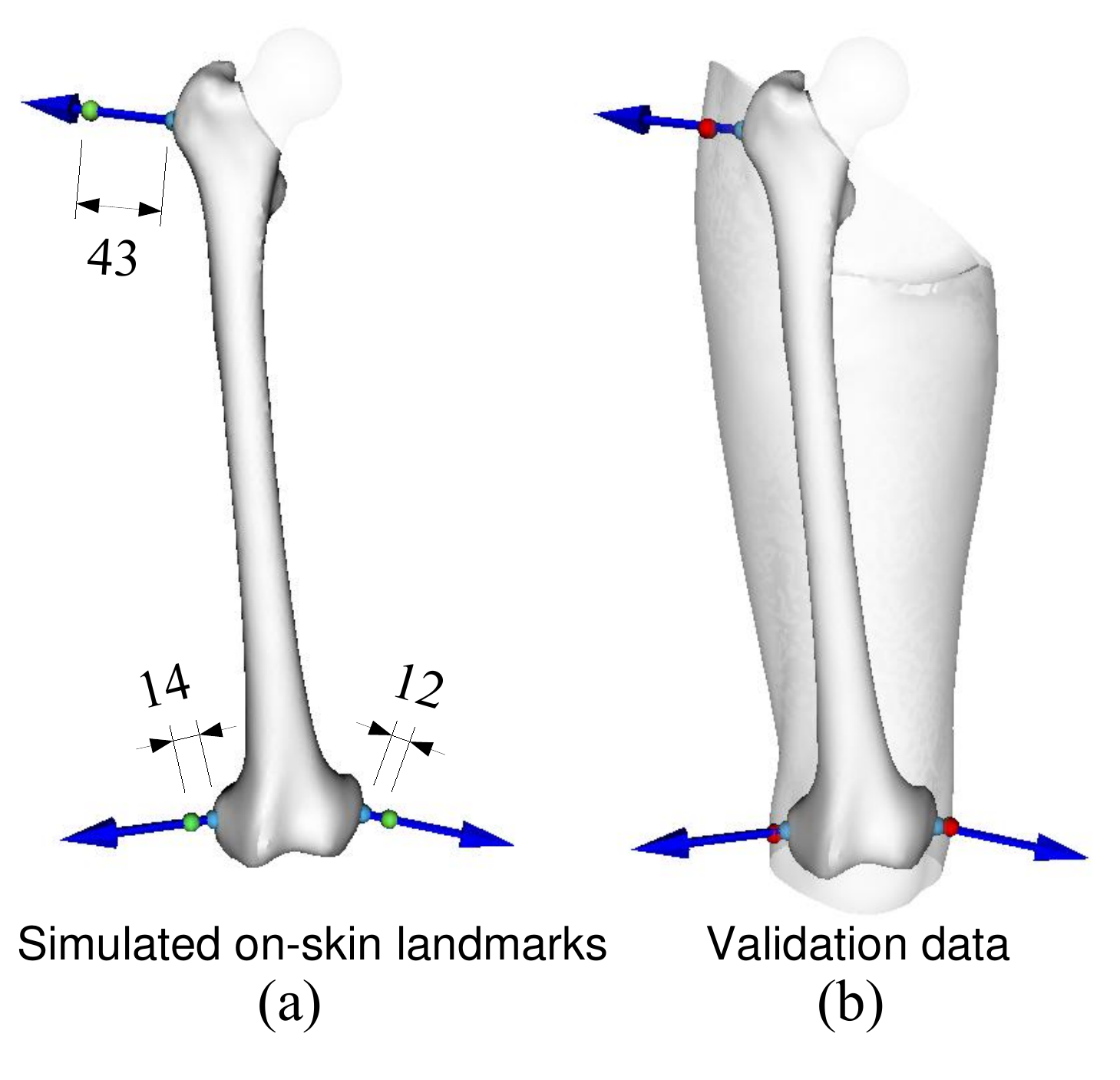} 
\caption{Two types of experiments were conducted to evaluate the use of skin data: (a) shows the selected bony landmarks (blue points) and their simulated on-skin correspondences (green points) in the surface normal direction, and (b) shows an example of our real validation data. The bony landmarks and on-skin landmarks are depicted by blue and red points, respectively.}  
\label{figure_onskin} 
\end{figure}

In the first experiment, the soft tissue thickness was simulated. An estimate of the distance between the bony landmarks and skin is required to approximate the error that can occur. The average soft tissue thickness over the desired bony landmarks was measured using CT scans of four cadavers and reported in Table \ref{three_landmarks}. 
To simulate the soft tissue, these positive biases were considered in the direction of the normal to the bone surface. The unseen femoral head and neck, and the mechanical axis were predicted under two different conditions, firstly when using bony landmarks, and secondly when using their simulated skin correspondences (see Fig. \ref{figure_onskin}-a).  

In the second experiment, instead of simulating the thickness of soft tissue, the distance between the real femur bone and thigh skin meshes were used.
The normals of the bone surface at the chosen bony landmarks were computed and the skin landmarks were selected as the closest vertices of skin mesh to the normals passing through the skin surface. The deformation field was computed as the collection of the vector fields from the selected bony landmarks in the SSM to their on-skin correspondences.
Figure \ref{figure_onskin}-b shows an example of our femur and thigh skin mesh data. The developed SSM on the Sicas train set that had a superior performance in the other experiments was used to perform this assessment. 

\begin{table} [b]
\footnotesize
\centering
\caption{The distances between the bony to skin landmarks (in mm).}
\label{three_landmarks}
\begin{tabular}{lcc}
\hline
{Landmark} & {Mean distance} & {Standard deviation} 
\tabularnewline 
\hline
Greater trochanter & 43 & 20\tabularnewline 
Medial epicondyle  &  14 & 6 \tabularnewline
Lateral epicondyle &  12 & 4  \tabularnewline
\hline
\end{tabular} 
\end{table}

\section{Results} \label{sec:experiments} 
\subsection{Correspondence analysis and SSM construction} 
The performance of the CPD-based fitting was computed and compared in terms of surface RMSE for each dataset (see Table \ref{rmse_fitting}). The mean error and the standard deviation were lower for the Sicas dataset.

\begin{figure} 
	\centering
	\includegraphics[width=8.1cm]{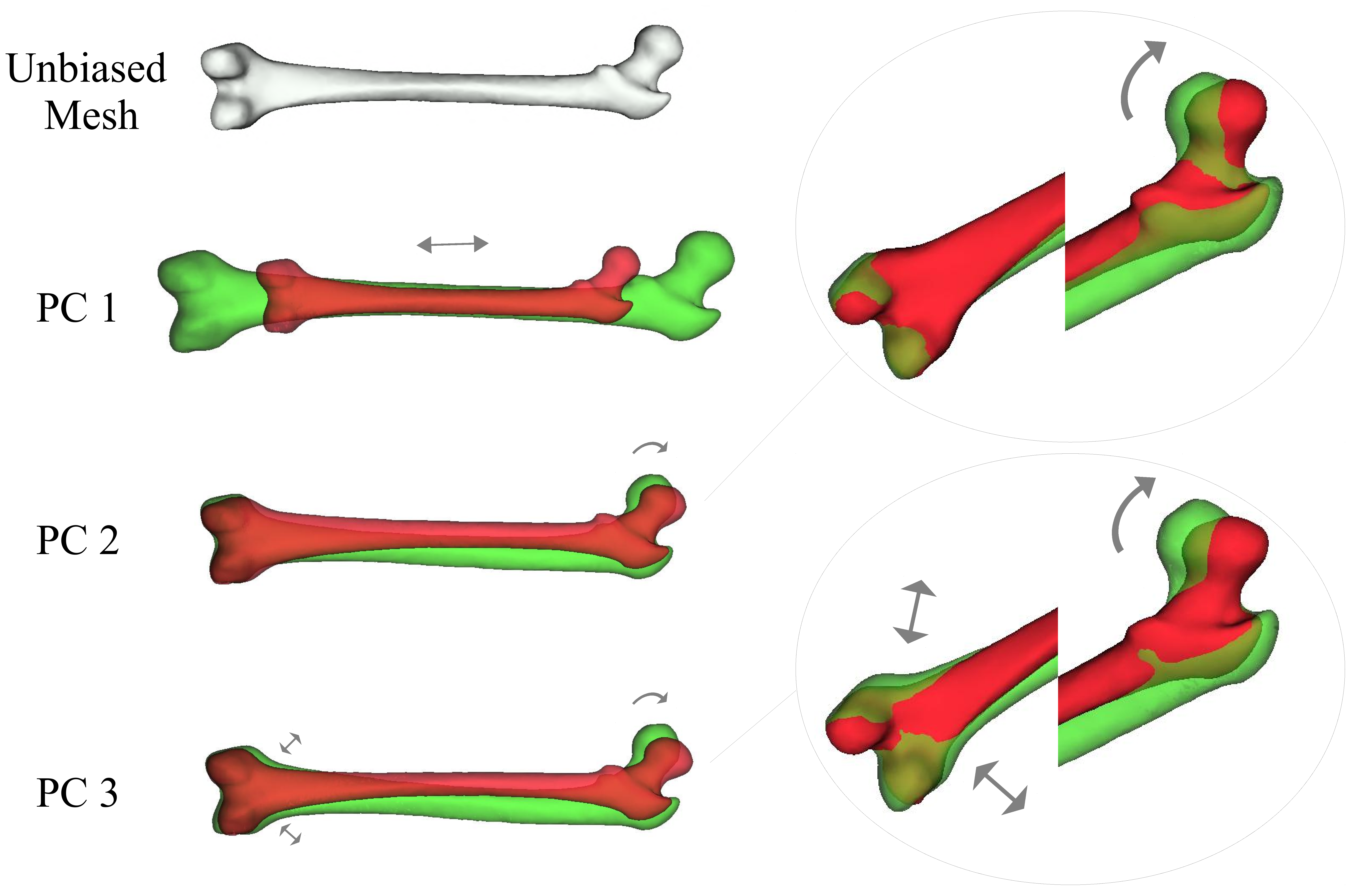}
	\caption{The computed SSM on Sicas data. The top femur in gray shows the unbiased mesh. The next rows show shape variation along with the first three Principal Components (PCs) of SSM. Represented in red and green are the variation of the mean femur by three times standard deviations along eigenmodes.} 
	\label{modes}
\end{figure}

Figure \ref{modes} illustrates three dominant modes of the developed SSM on the Sicas data. The first mode represents size variation. The second and third modes account for femoral neck angle, variation in anteversion angle, and shaft twist. The third mode also reflects the femur width.

\begin{figure*}
	\centering
	\begin{tabular}{ m{0.05cm} m{7.1cm} m{7.1cm}} 
		(a) & \includegraphics[height=6.1cm]{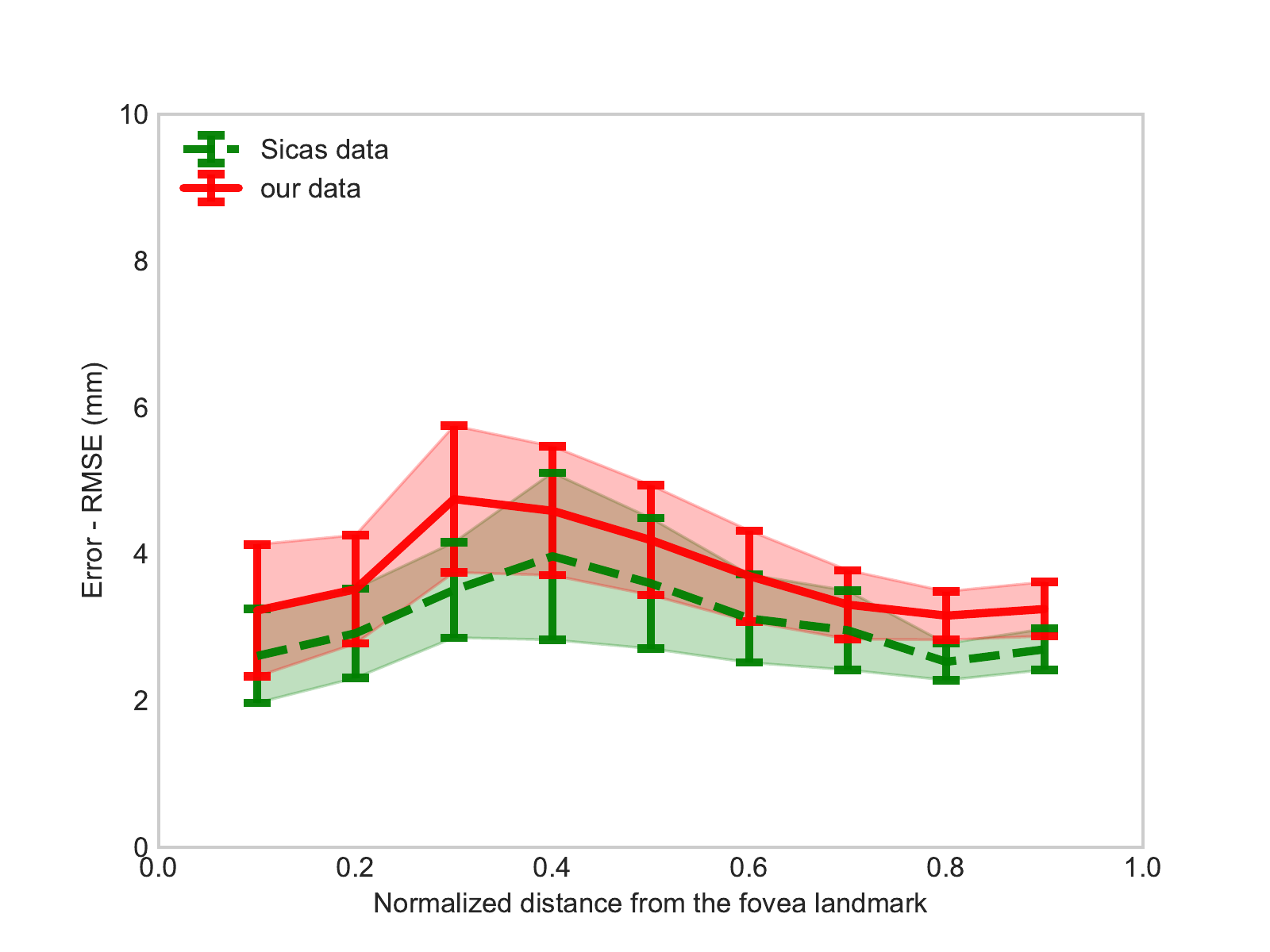} & \includegraphics[height=6.1cm]{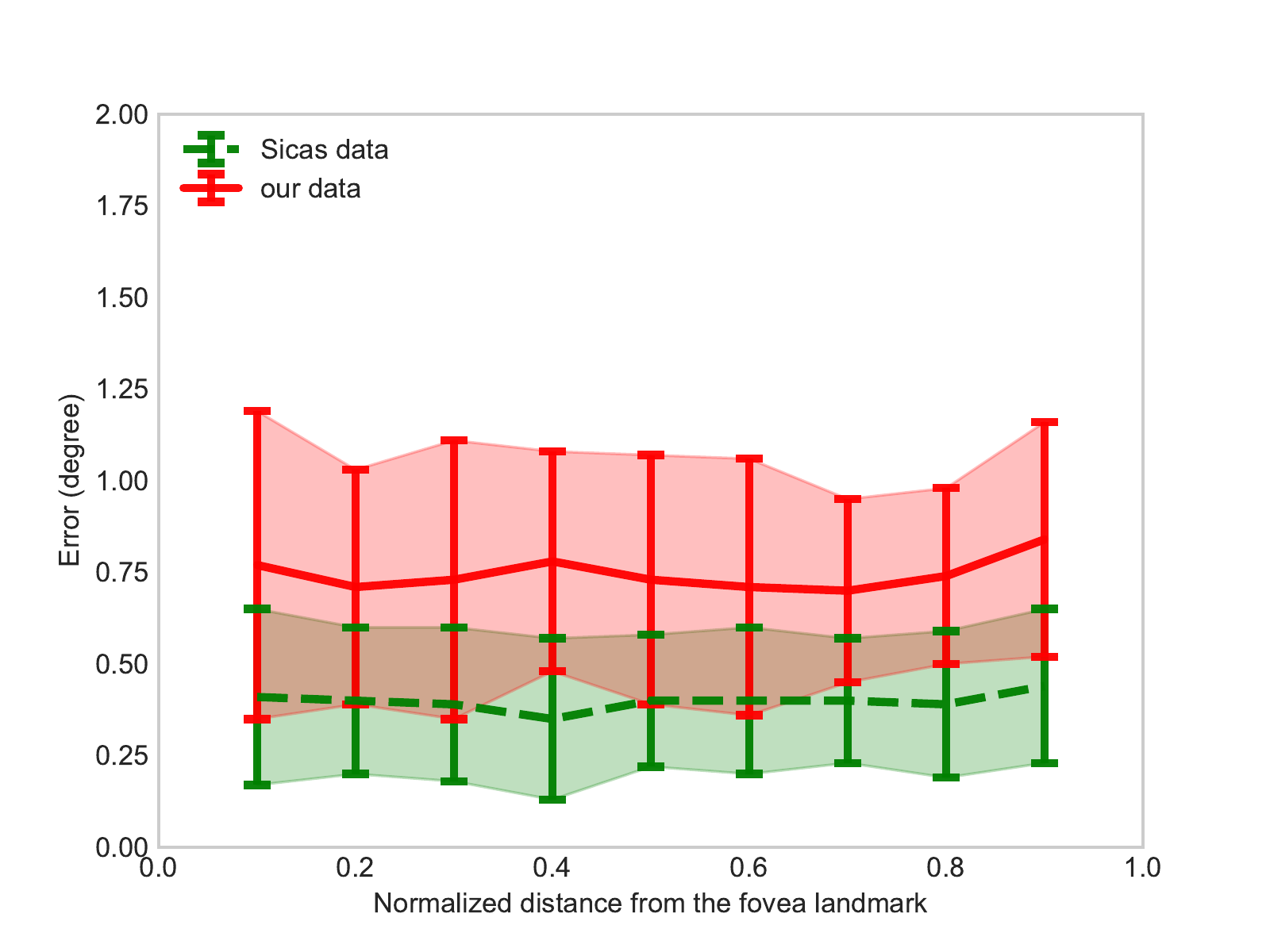} \tabularnewline
		(b) & \includegraphics[height=6.1cm]{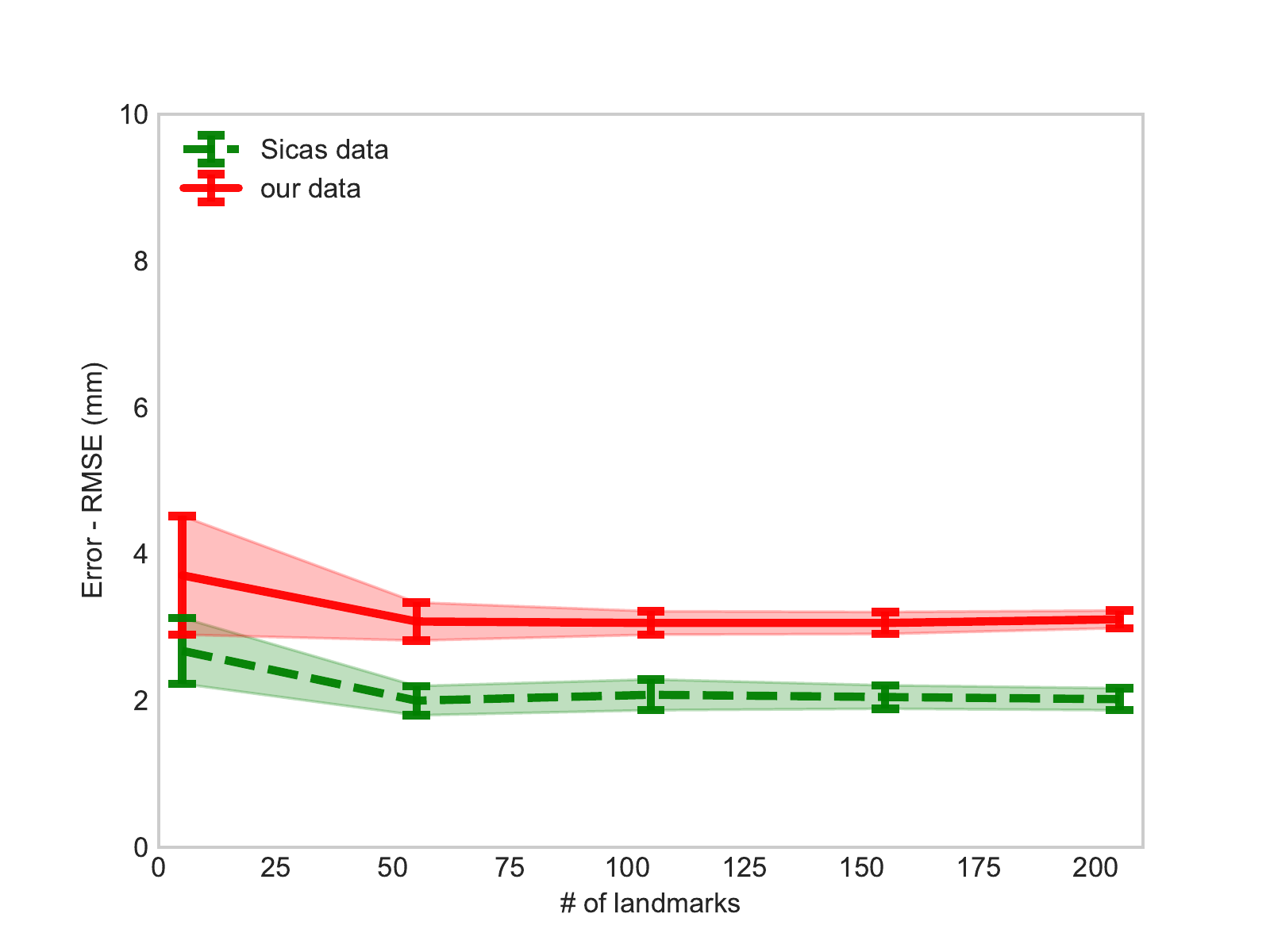} & \includegraphics[height=6.1cm]{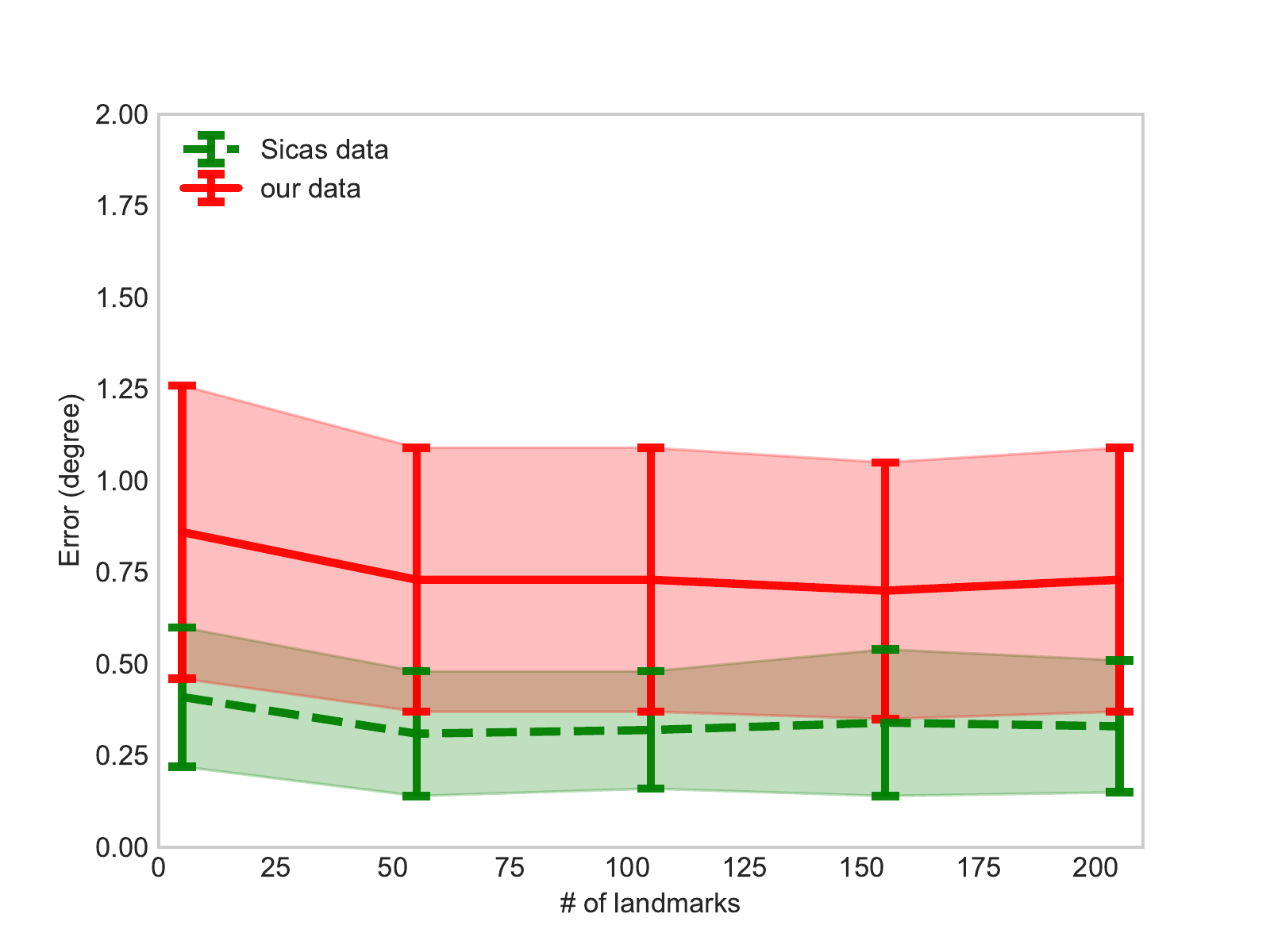} \tabularnewline
		(c) & \includegraphics[height=6.1cm]{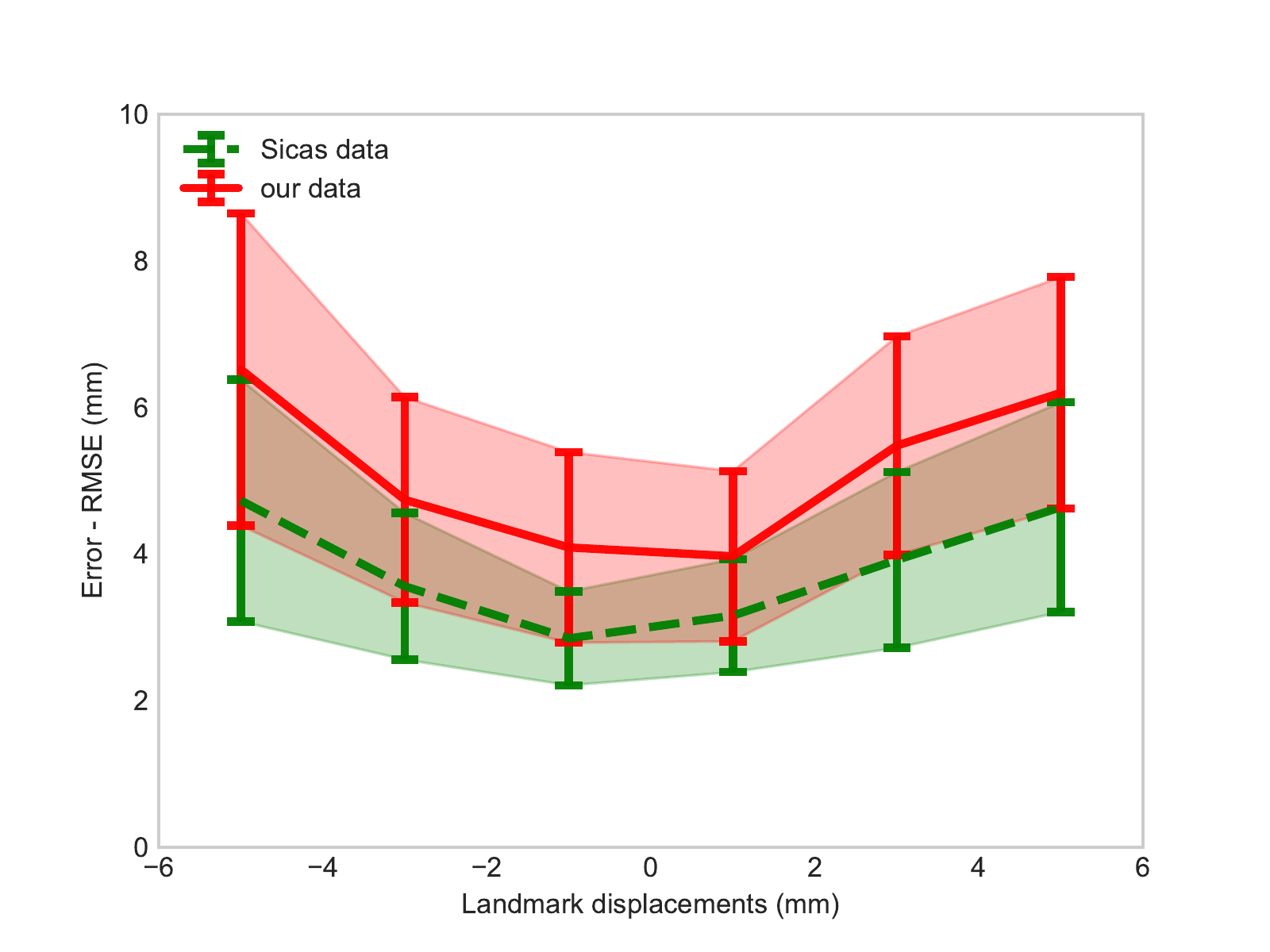} & \includegraphics[height=6.1cm]{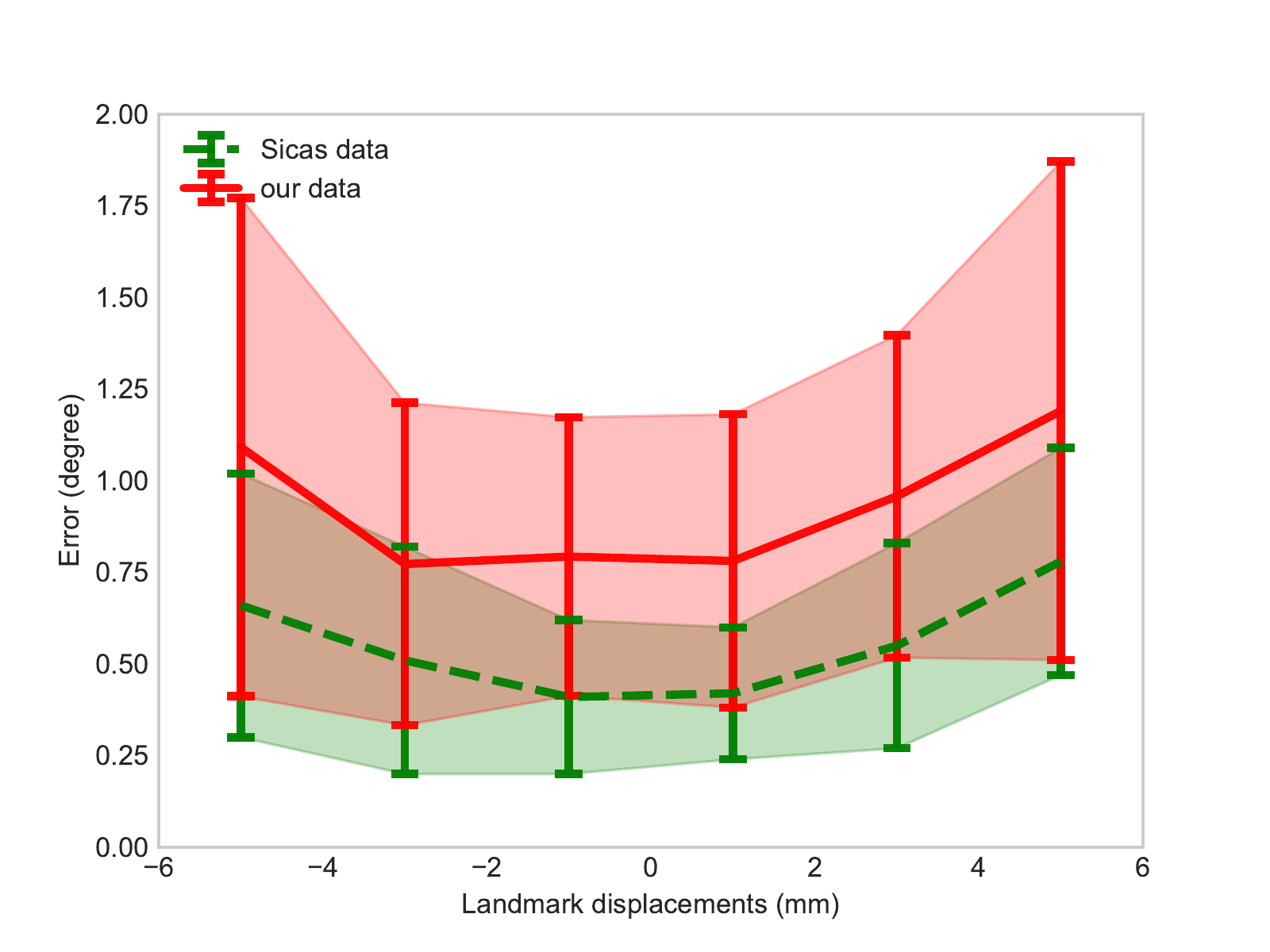}
	\end{tabular}
	\caption{The accuracy of predicted proximal femurs (a) when changing the distance of the landmarks from the fovea landmark, (b) when changing the number of the landmarks, and (c) when the landmarks are displaced. Left and right figures show the performance in terms of RMSE and mechanical axis angle deviation, respectively.}
	\label{figure_ring}
\end{figure*}

\begin{table} [b]
\footnotesize
\centering
\caption{The fitting error in the training sets (in mm).}
\label{rmse_fitting}
\begin{tabular}{lcc}
\hline
{Statics / Data} & {Our femur train set} & {Sicas train set} \tabularnewline 
\hline
Mean & 2.31 & 1.93 \tabularnewline 
Standard deviation & 0.93 & 0.15 \tabularnewline 
\hline
\end{tabular} 
\end{table}

\subsection{Performance with landmark configurations}
As shown in Fig. \ref{figure_ring}-a, we can observe that when the distance between landmarks and the fovea increased, RMSE also increase and then decreased. Specifically, the RMSE increases 2 mm when landmarks are located on the diaphysis.
According to Fig. \ref{figure_ring}-b, the mean RMSE decreases about 1 mm when the number of landmarks increased from 5 to 55. 
Finally, as it can be seen in Fig. \ref{figure_ring}-c, the RMSE increases between 2 and 3 mm when the landmark displacement increased by 5 mm. The mean angle deviation is never higher than 1.75$^\circ$ degrees.

\begin{figure*} [t!]
\centering
\begin{tabular}{c}
	\includegraphics[height=6.cm]{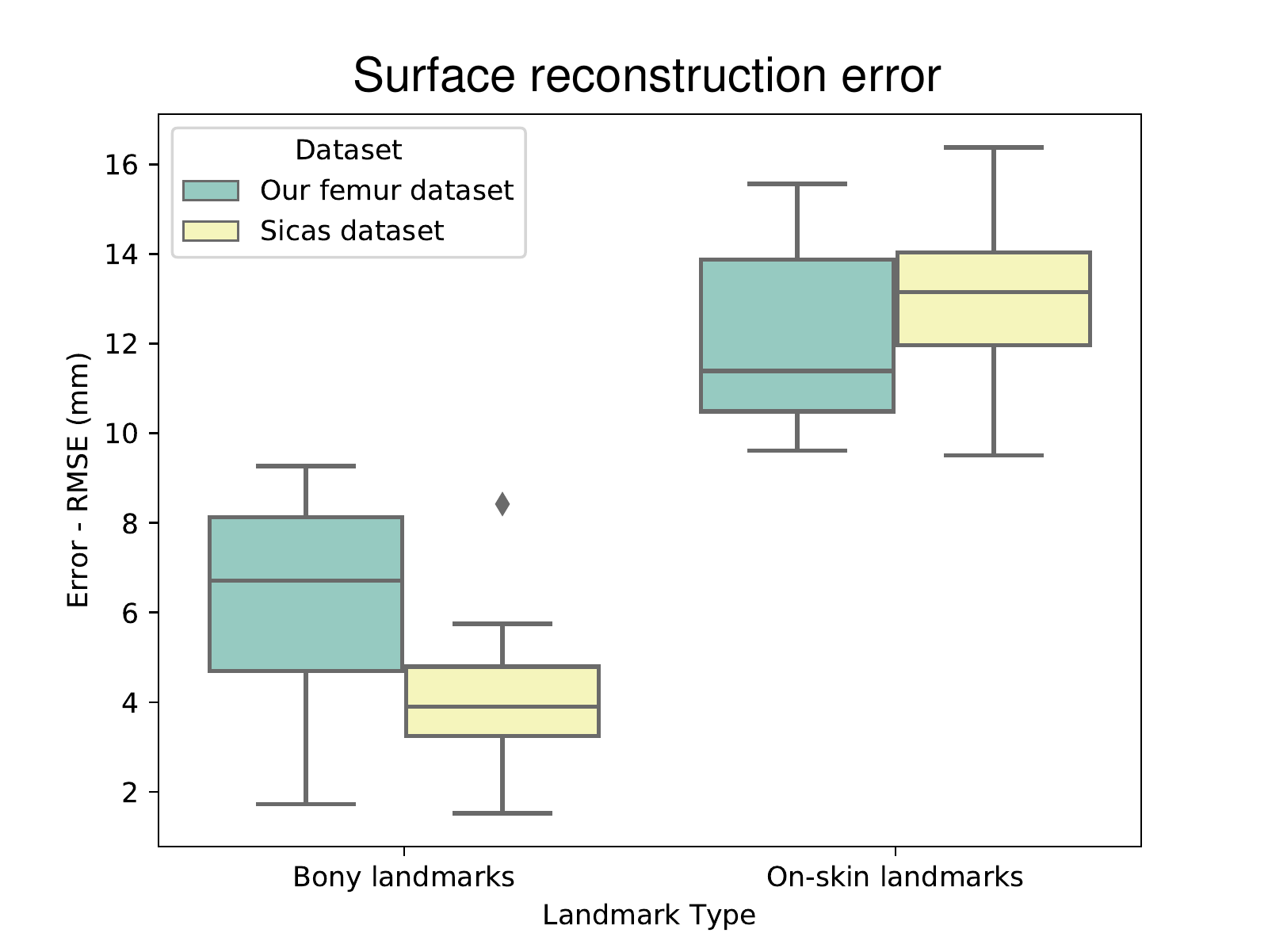} 
	\includegraphics[height=6.cm]{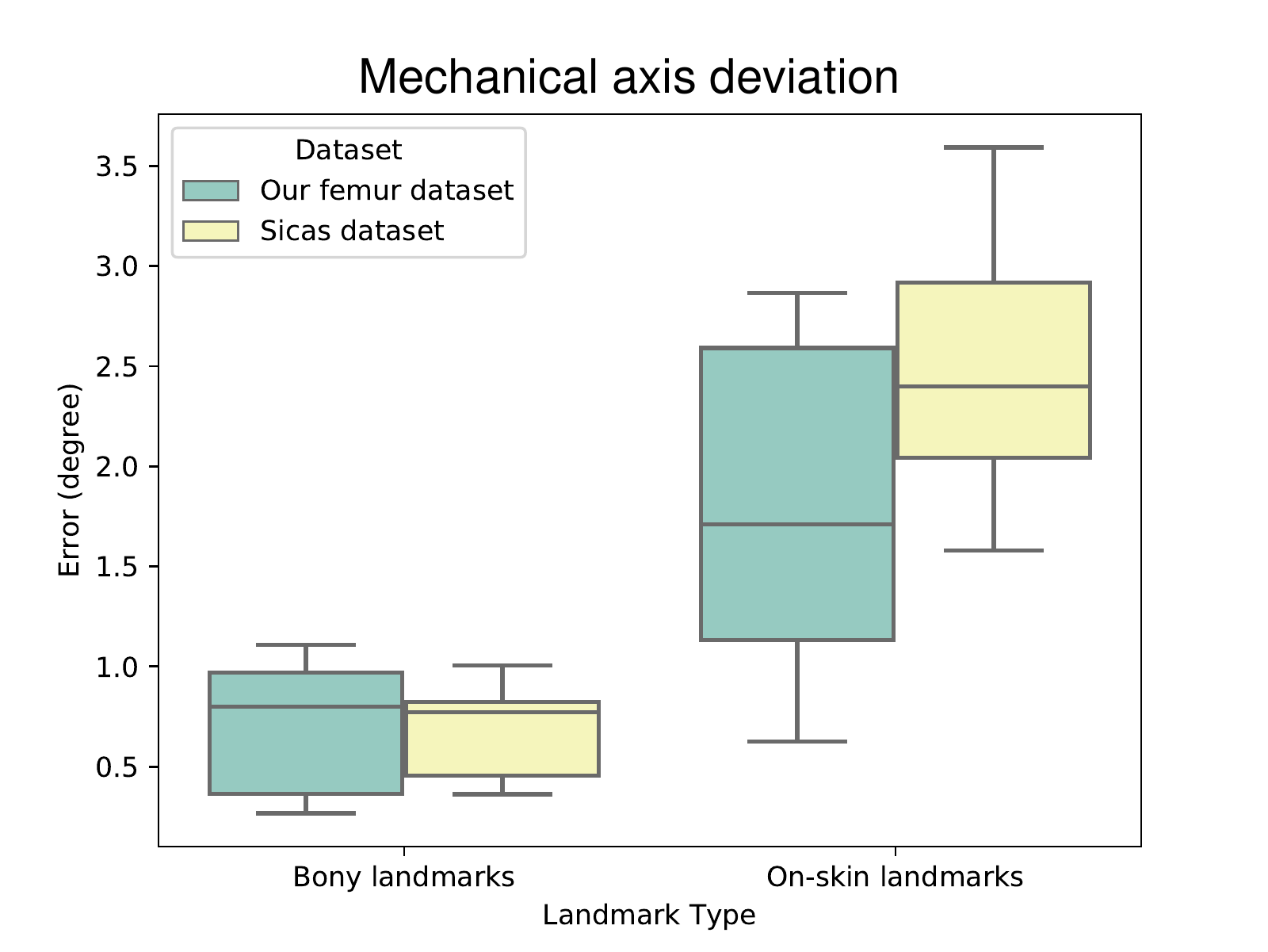} 
\end{tabular}
\caption{Evaluation of the predicted proximal femur using simulated on-skin landmarks. Left and right plots show the errors in terms of surface RMSE and the mechanical axis angle deviation error.}
\label{figure_skn}
\end{figure*}

\subsection{Exploring the use of skin landmarks} 
When simulated on-skin landmarks (see Fig. \ref{figure_onskin}-a) were used in replacement of bony landmarks, the surface RMS error was increased from less than 7 mm to about 12 mm (see Fig. \ref{figure_skn}). The median mechanical axis deviation error was increased from 0.75$^{\circ}$ degrees to 1.7$^{\circ}$ and 2.4$^{\circ}$ degrees in our dataset and SICAS dataset, respectively.  

\begin{figure} [t!]
\centering
\begin{tabular}{c}
\includegraphics[height=4.9cm]{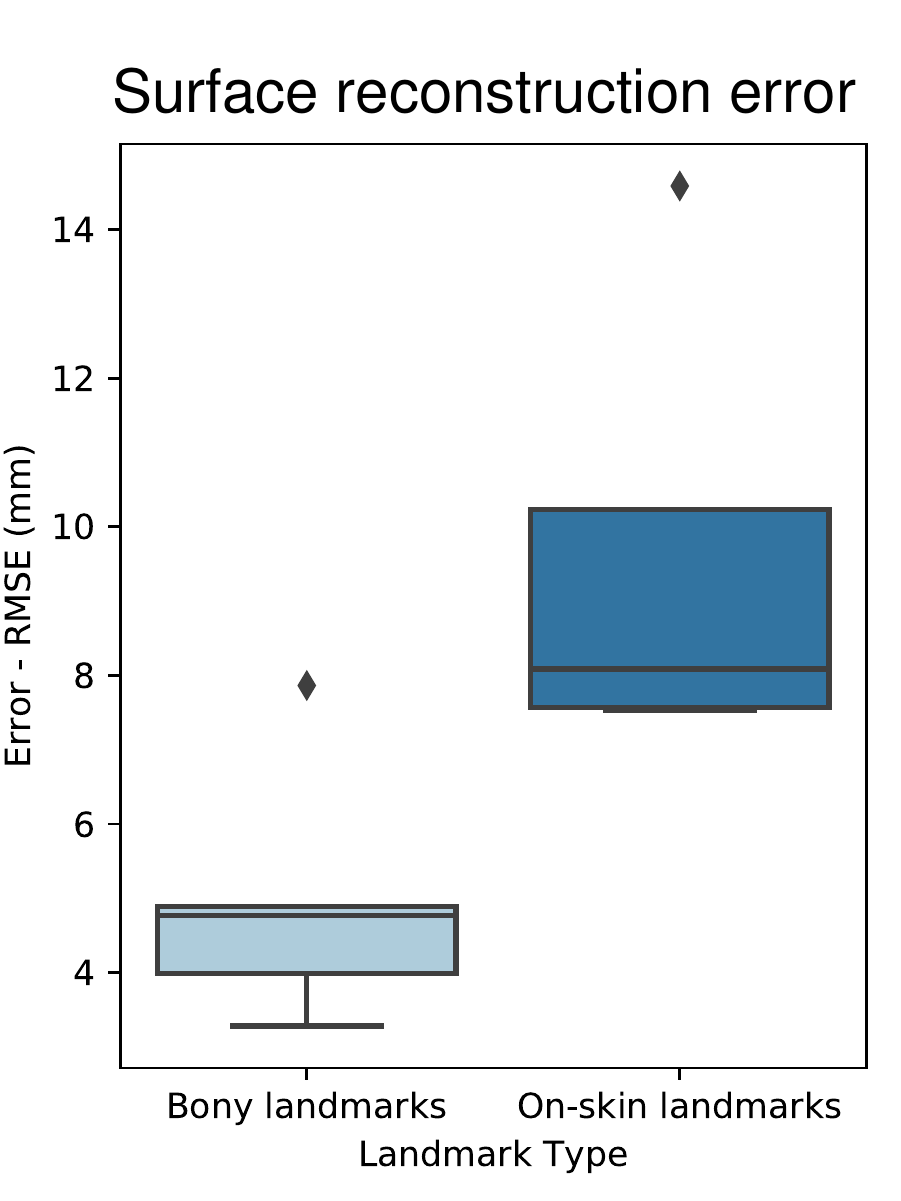} 
\includegraphics[height=4.9cm]{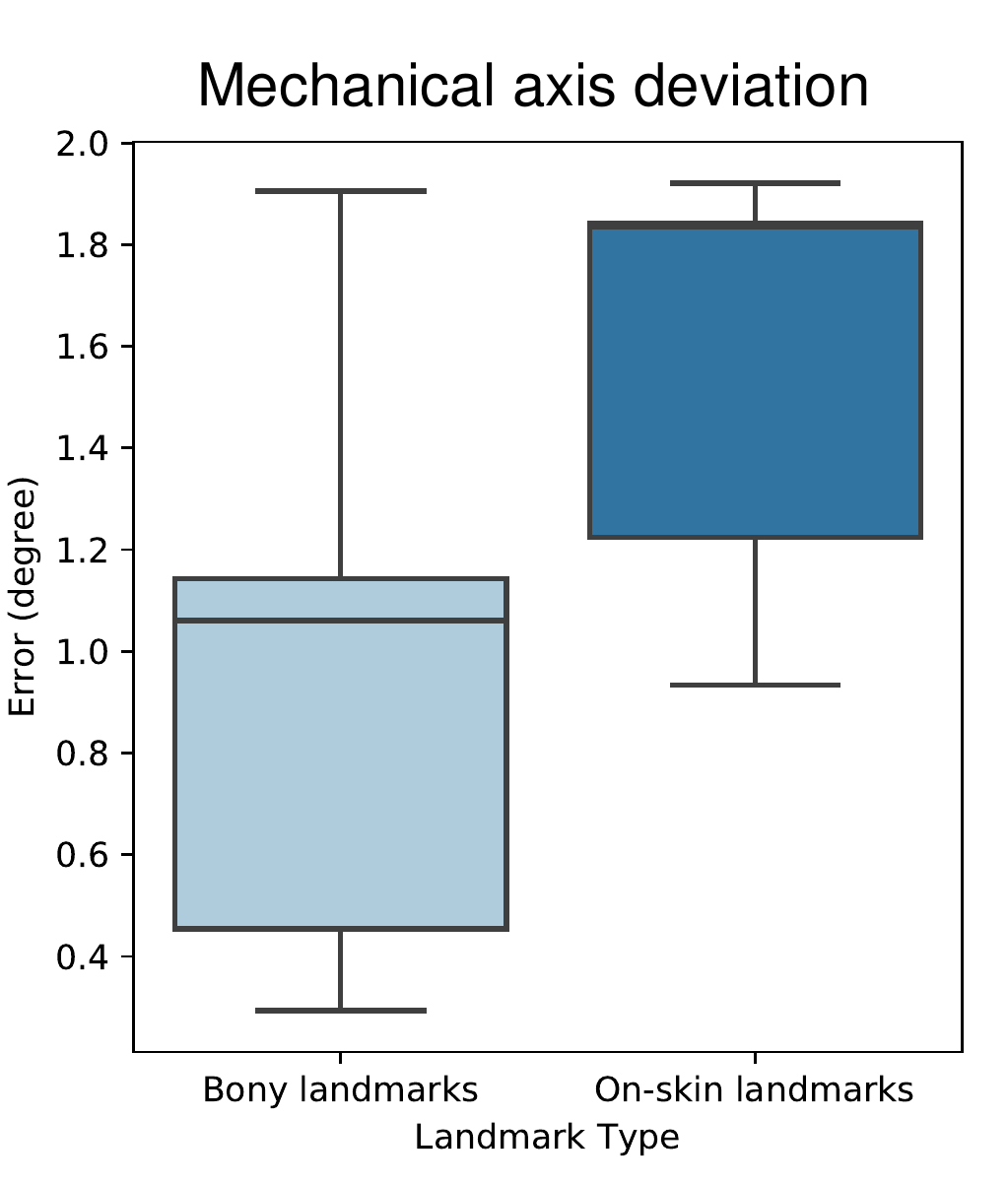}
\end{tabular}
\caption{The accuracy of predicted proximal femurs in `our femur bone and thigh skin mesh data' in terms of Left: RMSE (mm), and Right: mechanical axis angle deviation error.}
\label{reco_error}
\end{figure}

The surface RMS error using real skin landmark data (see Fig. \ref{figure_onskin}-b) was in the range of 7.5 to 10.5 mm, and the angle deviation error regarding the mechanical axis was less than $2 ^{\circ}$ degrees (see Fig. \ref{reco_error}).

\section{Discussion} \label{sec:discussion}
The objectives of our paper were to reconstruct the missing part of the femur shape and to retrieve the mechanical axis using a SSM and a few landmarks which can be acquired non-invasively. The experiments have been done first with landmarks that can be directly acquired on the bone surface during surgery, and then with landmarks acquired on the skin surface (non-invasive).

\subsection{Impact of landmark configurations} 
The inaccuracy in the landmark acquisition as well as the number of acquired landmarks can have a significant impact on the results.
It can be seen in Fig. \ref{figure_ring}-a that the acquisition of landmarks from areas with more curvature (\ie, landmarks located in the epiphyses) are more relevant as they better capture significant deformations of the shape. For instance, the x-axis at the values 0.1, 0.3-0.4, and 0.8 correspond to experiments with the near proximal, diaphysis, and distal regions, respectively.
Also, it was observed that increasing the number of landmarks leads to only a slightly more accurate shape reconstruction and the error converges quickly (no effect for more than 50 landmarks, as shown in Fig. \ref{figure_ring}-b). In particular, more than 5 landmarks only contribute marginally to the reliability of the deformation field and to the improvement of the prediction.
However, the inaccuracy regarding the acquired landmark positions can lead to a higher error regarding the prediction of the missing part (Fig. \ref{figure_ring}-c).
These findings highlight thus the importance to have a good configuration of landmarks to improve the reconstruction performance. But this configuration has to take also into account the clinical requirements, \ie, landmarks easily recognizable either during surgery or percutaneously.

\subsection{Impact of two different datasets} 
One of the questions we faced in this study was, how to make the best use of two datasets (different in size, length, and distribution; please see Section \ref{sec:specimens}). If the goal was to create a single model (and to estimate its generalization error), then it made sense to first combine the data and next do the rest of the analysis. However, our goal was to analyze a parameter (the landmark configuration) in the constructed model and to verify the reliability of our analysis, therefore we found it rational to perform an individual analysis on each dataset and then compare the findings. The bias and variance of results in several iterations on two datasets were reported in Fig. \ref{figure_ring}.
The same pipeline was used on both datasets. It was noticed that the obtained results in the Sicas dataset were more accurate, about 0.5 -- 2 mm surface RMSE, in most of the experiments (see Fig. \ref{figure_ring}). The higher predictive quality of the model developed on the Sicas data was because, probably, of the larger training set.

\subsection{Comparison with related works} 
Zhang and Besier \cite{zhang2017accuracy} obtained an error of 19.1 mm and 1.8 mm respectively using linear scaling and SSM.
Barratt et al. \cite{barratt2008instantiation} reported a mean surface error between 3 -- 4 mm for the SSM registration to the femur and the pelvis. 
In comparison with their works and using only three bony landmarks the RMSE was lower than 5 mm.
The obtained accuracy using on-skin landmarks was between 7.5 -- 10.5 mm.
Although surface reconstruction using SSM and a few on-skin landmarks shows higher accuracy than the linear scaling method in Zhang and Besier \cite{zhang2017accuracy}, the accuracy is still lower than when SSM and a large number of bony landmarks were used.
The deviation regarding the mechanical axis was always inferior to $2 ^{\circ}$ degrees for bony landmarks, and $3.5 ^{\circ}$ degrees for on-skin landmarks. Although the surface reconstruction accuracy is probably not enough for several surgical applications, the results regarding the mechanical axis are very encouraging and could potentially open a very interesting perspective for the analysis of the lower limb alignment either for orthopedics with direct access to the landmarks or for functional rehabilitation with the on-skin version.	
In comparison, the accepted accuracy in total knee arthroplasty is $3 ^{\circ}$ degrees \cite{rahm2017postoperative}, and the accepted accuracy regarding the lower limb alignment is $2 ^{\circ}$ degrees for high tibial osteotomy, \ie, $3 ^{\circ}$--$5 ^{\circ}$ degrees of valgus in the mechanical axis is considered to be an optimal correction \cite{sabzevari2016high}.

\subsection{Limitations of the study and future scope} 
In this paper, the experiments have been carried out on a single train-test split on two independent datasets. The best practice would be performing the cross-validation protocol which gives a superior approximation of generalization.
In fact, the implementation of two cross-validations is required (on our dataset and sicas dataset). However, because of the high number of parameters in our study, launching and running cross-validation would be very long therefore it was avoided.
Two SSMs were developed separately on each train set. The modes of variations in the SSMs (see the Sicas-trained SSM in Fig. \ref{modes}) were in agreement with the previous literature studies \cite{zhang2014anatomical, zhang2016predictive, gaffney2019statistical}.
Although our objective was not to analyze the quality of SSMs, there could be a further analysis looking at quantifying the intrinsic quality of the obtained SSMs.
It also should be mentioned that a cadaveric dataset was used in this study. The cadavers are often frozen that can suppress (although not significantly) the soft-tissue deformations.
Finally, we did not assess in this study the impact of the soft-tissue deformation regarding the on-skin landmark-based deformation. But, we think that our testing scenario was maybe the worst case since by applying a small force on the skin during the palpation of the landmarks, we could probably improve the results, because of the lower distance between the on-skin and on-bone landmarks. However, this assumption has to be of course further analyzed.

To conclude, in this study we explored how the quality of landmark correspondences impacts the reconstruction of the missing proximal femur surface. In addition, we investigated the effectiveness of using skin landmarks instead of bony landmarks.
Although, the obtained surface prediction accuracy using skin landmarks is insufficient for many clinical applications that require a high accuracy (\eg, designing personalized implants), the result of the non-invasive determination of the mechanical axis is very encouraging and can be of particular interest for studying the lower limb alignment in functional rehabilitation.
Interesting future directions could be to integrate both skin and bone information into a combined SSM and during the reconstruction procedure, and to evaluate more clinical parameters that are of concern from the clinical point of view.

\section*{Acknowledgments}
The authors would like to thank the editor and referees for their helpful comments and suggestions.
This work is supported by ``Investissements d'Avenir programme Labex CAMI'' under the reference ANR-11-LABX-0004. 

\bibliographystyle{elsarticle-num} 
\bibliography{references}

\end{document}